\documentstyle[graphicx,prl,multicol,amsmath,aps,epsfig]{revtex}

\begin{document}

\draft

\sf                      %

\title{Low-frequency flux noise and visualization of
vortices in a $\rm\bf YBa_2Cu_3O_7$ dc SQUID washer
with an integrated input coil}

\author{R. Straub, S. Keil, R. Kleiner}
\address{Universit\"{a}t T\"{u}bingen, Experimentalphysik II, D-72076 T\"{u}bingen, Germany}
\author{\renewcommand{\thefootnote}{\alph{footnote})}
D. Koelle\footnote{e-mail: koelle@ph2.uni-koeln.de}}
\address{II. Physikalisches Institut, Universit\"{a}t zu K\"{o}ln, D-50937 K\"{o}ln, Germany}
\maketitle

\begin{abstract}
We used low-temperature scanning electron microscopy
(LTSEM) for imaging quantized magnetic flux (vortices)
in direct current (dc) superconducting quantum
interference devices (SQUIDs) with approximately $\rm
1\,\mu m$ spatial resolution at temperature $T=77\,\rm K$
in a controllable magnetic field up to $\rm 20\,\mu T$.
We demonstrate that LTSEM allows to image the spatial
distribution of vortices in a
$\rm YBa_2Cu_3O_7/SrTiO_3/YBa_2Cu_3O_7$
multilayer thin-film structure consisting of a dc SQUID
washer with an integrated input coil on top.
Simultaneously, we are able to measure the low-frequency
noise of the sample under test, which allows to
correlate the spatial distribution of vortices with
low-frequency noise in the SQUID.
\end{abstract}

\vspace*{5mm}

%
%


\vspace*{-90mm}\noindent to be published in Applied Physics
Letters
\vspace*{82mm}

\begin{multicols}{2}

In recent years impressive progress was achieved
towards the development of sensitive superconducting
quantum interference devices (SQUIDs) based on high
transition temperature ($T_c$) superconductors
\cite{Koelle99}. For high-$T_c$ SQUID magnetometers made
from epitaxially grown thin films of $\rm
YBa_2Cu_3O_7$ (YBCO) and $\rm SrTiO_3$ (STO)
insulating layers an rms magnetic field resolution
$S_B^{1/2}<\rm 10\,fT/\sqrt{Hz}$ in
the white noise limit at temperature $T=77\,\rm K$
was reported for the best devices
made so far. 
More typical
values are some tens of $\rm fT/\sqrt{Hz}$. Such levels
of white noise are adequate for most applications of SQUIDs.
Unfortunately, most devices do not achieve these low
noise levels at low frequency $f$, say below 10-100 Hz,
which makes them less suitable for many applications
\cite{Koelle99}.

The main source for the excess low-frequency noise,
which typically scales as $S_B\propto 1/f$, is the
thermally activated vortex motion in the high-$T_c$
thin films. This problem becomes even more serious for devices operated in
magnetically unshielded environment and cooled in the
earth's magnetic field, due to an increased density of
fluctuating vortices.
It is known that the low-frequency flux noise is correlated
with the quality of the high-$T_c$ films
\cite{Ferrari94}, which in turn is affected by the
presence of a variety of defects in these films.
However, a
detailed understanding of the interplay between
microstructure and noise properties of high-$T_c$ thin
films is still lacking. This hinders further improvement
of the low-frequency performance of high-$T_c$ SQUIDs
and magnetometers. Furthermore, the geometry of the
micropatterned devices and the patterning process
itself may significantly affect their noise properties.
Hence, a number of specific locations within a device
may represent possible noise sources.

It is obvious, that the low-frequency noise
of high-$T_c$ SQUIDs and magnetometers depends on local
properties determined by defects and geometry. Noise
measurements, however, give only spatially integrated
information on the device properties, which makes it
difficult to locate noise sources. In this work we
utilize a technique for the visualization of vortices
in SQUIDs to locate possible noise sources. This
technique is based on a new mode of operation of
low-temperature scanning electron microscopy (LTSEM),
which has been used previously for imaging the spatial
distribution of $T_c$ or critical current density $j_c$
in superconducting thin films and Josephson junctions
and of Josephson vortices in long junctions
\cite{Gross94}. In contrast, the LTSEM technique
presented here exploits the high sensitivity of SQUIDs
to magnetic flux directly. That is, we detect
electron-beam-induced changes $\delta\Phi$ of the
magnetic flux coupled to the SQUID loop, where the
SQUID itself is the device under test. Simultaneously,
we measure the low-frequency flux noise of the SQUIDs.
This enables us to obtain both images of
the spatial distribution of vortices and local
information on noise properties of the devices under
investigation.

We mount our high-$T_c$ SQUIDs on a magnetically
shielded, liquid nitrogen cooled cryostage of an SEM as
described in detail in \cite{Gerber97}. The dc SQUIDs
are read out using a standard flux-locked loop (FLL)
with 3.125 kHz bias current reversal to eliminate 1/$f$
noise due to fluctuations in the critical current $I_c$
of the Josephson junctions. The YBCO dc SQUIDs we have
investigated so far by LTSEM yield typical levels of rms
spectral density of flux noise
$S_\Phi^{1/2}=10-20\,\mu\Phi_0/\sqrt{\rm Hz}$ in the
white noise limit at $T=77\,\rm K$. For the spatially
resolved measurements, the e-beam is used as a local
perturbation which induces an increase in temperature
$\delta T(x-x_0,y-y_0)$ on the sample surface (in the
$x$-$y$-plane) centered around the beam spot position
$(x_0,y_0)$. The length scale for the spatial decay of
the thermal perturbation is set by the beam-electron
range $R\approx 0.5\,\rm\mu m$ for a typical beam
voltage $V_b=\rm 10\,kV$ \cite{Gross94}. This gives a
maximum increase in beam induced temperature $\Delta
T\approx\rm 1\,K$ at $(x_0,y_0)$ for a typical beam
current $I_b=1\,\rm nA$. So-called $\delta\Phi$-images
are obtained by recording the e-beam induced flux change
$\delta \Phi (x_0,y_0)$ as a function of the e-beam
coordinates $(x_0,y_0)$. To improve the signal to noise
ratio, we use a beam-blanking unit operating at
typically 2.6\,kHz and the  output signal of the FLL,
i.e. the e-beam induced flux change in the SQUID, is
lock-in detected. This signal controls the brightness of
a video screen as a function of $(x_0,y_0)$.
Additionally, the time trace or the power spectrum of
the FLL output signal can be recorded by a signal
analyzer.

\begin{figure}[t]
\center{\includegraphics [width=0.95\columnwidth,clip]
{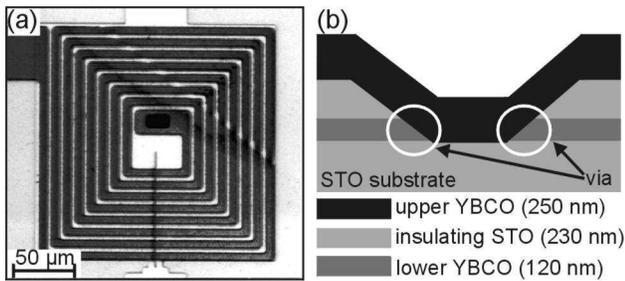}}
\caption[]{
(a): Optical micrograph of square washer in bottom YBCO (white)
with integrated input coil in top YBCO (dark grey).
(b): Sketch of cross section of the superconducting interconnect
(via). The white circles indicate the location of the via edge,
where top and bottom YBCO form a superconducting interconnect.
\label{figure1}}
\end{figure}

All measurements presented in this work were performed
at $T=77\,\rm K$ on one integrated Ketchen-type magnetometer based on a
YBCO/STO/YBCO multilayer structure \cite{Ludwig95}
with
a dc SQUID square washer in the bottom YBCO and a
spiral input coil in the top YBCO film, as shown in
Fig.~\ref{figure1}(a). The SQUID hole is a narrow slit,
visible as a dark vertical line in Fig.~\ref{figure1}(a),
running from the center to the bottom end of the washer.
At the innermost turn of the input coil a via connects
both YBCO layers along the edge of the black
rectangular area visible in the center of
Fig.~\ref{figure1}(a). Figure \ref{figure1}(b) shows a
cross section of the via. The pickup loop (not shown)
had been cut for investigation of the transport
properties of the input coil and the via
\cite{Gerber96}. For details regarding device
fabrication and geometry see
\cite{Ludwig95,Gerber96}.

A variety of e-beam induced changes in the output signal
of various YBCO dc SQUIDs has been observed and
described in more detail in \cite{Keil99,Koelle00}. The
mechanism for the imaging of vortices is also explained
in \cite{Keil99,Koelle00} and can be briefly described
as follows: The e-beam induced local increase in
temperature induces a local increase in the London
penetration depth $\lambda_L$. Hence, the screening
currents circulating around a vortex are spatially
extended due to e-beam irradiation. If the e-beam is
scanned across a vortex this vortex will be dragged
along with the beam, i.e. it will be displaced by some
distance $\delta r$, if the beam spot is within the
radial distance $R$ from the vortex. This displacement will
couple a positive (negative) flux change $\delta\Phi$
to the SQUID if the vortex is moved away from (towards)
the SQUID hole.

\begin{figure}[tbh]
\center{\includegraphics [width=0.95\columnwidth,clip]
{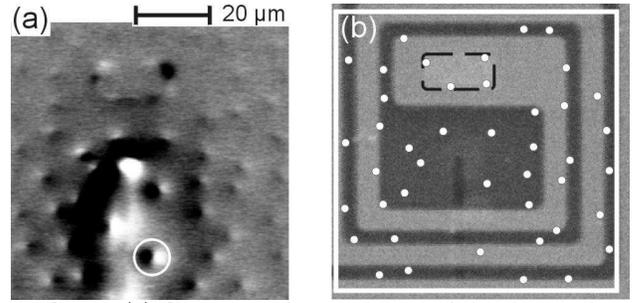}}
\caption{
(a) $\delta\Phi$-image of central part of the device shown in
Fig.~\ref{figure1} ($B_0=20\,\rm\mu T$);
the white circle indicates position of a particular
vortex mentioned in the discussion of
Fig.~\ref{figure3};
(b): SEM micrograph of central part of the device with the two
innermost turns of the input coil (light grey); the black dashed
frame indicates the position of the via. The white frame
corresponds to the frame of the $\delta\Phi$-image shown in (a).
The white dots indicate positions of vortices as imaged in (a).
\label{figure2}}
\end{figure}

Fig.~\ref{figure2}(a) shows an example of a
$\delta\Phi$-image with approximately 50 vortices
appearing as pairs of positive (bright) and negative
(dark) signals. This image was taken from the center of
the device after cooling in a static magnetic field
$B_0=20\,\rm\mu T$ from above $T_c$ to $T=77K$. The
maximum displacement $\Delta r$ of a vortex is of the
order of the change in $\lambda_L$ (typically $\sim$ 20
nm in our experiments at 77\,K). This maximum
displacement $\Delta r$ induces a maximum signal
$\Delta\Phi=(\partial\Phi/\partial r)\Delta r\equiv\phi_r(r)\Delta r$,
with $r$ being the radial distance
of the vortex from the SQUID hole. As $\Delta r$ is
fixed for fixed $I_b$, $V_b$ and $T$, the signal
$\Delta\Phi$ is a direct measure of the coupling strength
$\phi_r$ of an imaged vortex. Together with $\phi_r$,
the vortex signal $\Delta\Phi$ decreases rapidly with
increasing $r$ \cite{Ferrari91,Humphreys99}. Hence, only
vortices within a radial distance of some 10 $\mu$m give
strong signals. An important advantage of our imaging
technique, however, is the fact that the spectral
density of
flux noise $S_\Phi$ induced by motion of
a single vortex is given by
$S_\Phi=S_r\cdot\phi_r^2$, with $S_r$ being
the spectral density of radial motion of this vortex.
Therefore, measuring $\phi_r$ directly gives important
information on possible noise sources in the
investigated devices. This is particularly important
for complex structures, as the one described in this
letter, for which calculation of $\phi_r(x,y)$ is not
straightforward.

The correlation of $\delta\Phi$-images with surface
micrographs provides information on the position of
individual vortices. For this purpose we marked the
position of vortices obtained from Fig.~\ref{figure2}(a) as
white dots in Fig.~\ref{figure2}(b) (surface
micrograph). We find that most vortices are located
between the turns of the input coil and hence are
penetrating only the lower YBCO film. This observation
implies, that the 7 $\mu$m wide input coil is
narrow enough to prevent entry of most vortices
\cite{Dantsker97}, at least
up to $B_0=20\,\rm\mu T$.
We observe a tendency of the vortices being located
close to the edge of the input coil. We have not
clarified yet whether this is due to preferential
pinning at the edges. Alternatively, this may be due to
a slightly higher $T_c$ of the upper YBCO as compared
to the bottom YBCO film\cite{Gerber96}:
Screening currents induced upon cooling the upper YBCO
film through $T_c$ in a field $B_0$ create an inhomogeneous flux
density distribution in the bottom YBCO film with enhanced
flux density near the edges of the input coil, which
may favor nucleation of vortices there. Possibly for
similar reasons, we always observe vortices to be
located right at the via edge which is indicated as a
dashed frame in Fig.~\ref{figure2}(b). As shown
previously \cite{Gerber96}, the upper YBCO film inside this frame has
the highest $T_c$ of this structure. In
Fig.~\ref{figure2} we see four vortices pinned at
the via edge, three of them exactly in the corners of
the rectangle. We note that these four vortices give
signals which are spatially more extended as compared to
other vortices. This could be due to an
increased value of $\lambda_L$ at the via edge, related
to reduced screening, in accordance with reduced
critical current density measured for our vias
\cite{Ludwig95}. However, due to the complex geometry
of the via edge, which involves c-axis transport,
interpretation of these signals is not straightforward.
In addition these vortex signals are also stronger than
from other vortices at similar $r$. It is not clear
yet, whether this indicates a higher mobility of these
vortices, which in turn would make them strong
candidates for prominent low-frequency noise sources.

\begin{figure}[tbh]
\center{\includegraphics [width=0.86\columnwidth]{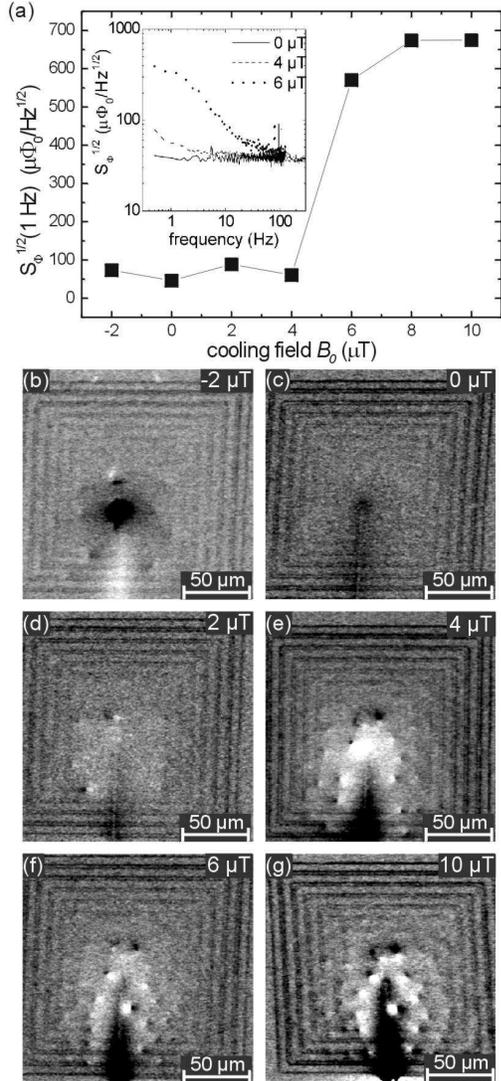}}
\caption{
Low-frequency noise and
$\delta\Phi$-images: (a) rms spectral density
of flux noise $S_\Phi^{1/2}$ at $f=1\rm\,Hz$ vs.~cooling
field $B_0$. The inset shows $S_\Phi^{1/2}(f)$ for selected
values of $B_0$; (b)-(g): $\delta\Phi$-images,
taken directly before or after measuring $S_\Phi$.
\label{figure3}}
\end{figure}
To correlate the frequently observed increase of
low-frequency flux noise for increasing cooling field
$B_0$ with the spatial distribution of vortices we took
a series of $\delta\Phi$-images from our device and
measured its spectral density of flux noise directly
before or after taking the image
for a series of values of $B_0$ in the range $-2\,\rm\mu T\le B_0\le 10\,\rm\mu T$.
Fig.~\ref{figure3}(a) shows the flux noise vs.~cooling
field measured at $f=1\,\rm Hz$ with a pronounced
increase in $S_\Phi(1\,\rm Hz)$ above $B_0=4\,\rm\mu
T$.
For $B_0=0\,\rm\mu T$ the noise spectrum shown in the
inset of Fig.~\ref{figure3}(a) is white down to below
$f=1\,\rm Hz$. This corresponds to the observation of
no vortices in Fig.~\ref{figure3}(c) and the lowest
value for $S_\Phi(1\,\rm Hz)$ in Fig.~\ref{figure3}(a). For
$B_0=-2\,\rm\mu T$ a few vortices appear
[c.f.~Fig.~\ref{figure3}(b)] which induce a slight
increase in low frequency noise [c.f.~Fig.~\ref{figure3}(a)].
The situation is similar for
$B_0=2\,\rm\mu T$, with opposite polarity
of the
vortices as shown in Fig.~\ref{figure3}(d). Increasing
$B_0$ to
$\rm 4\,\mu T$
increases the number of vortices [c.f.~Fig.~\ref{figure3}(e)], however, without a significant change in
the low-frequency noise [c.f.~Fig.~\ref{figure3}(a) and
inset].

At $B_0=6\,\rm\mu T$ a jump in the noise power by
almost two orders of magnitude occurs, which comes
along with the appearance of one particular vortex
located [c.f.~Fig.~\ref{figure2}(a)] at the second innermost turn, close to the
place where the turn crosses over the slit in the SQUID
washer. This location has previously been identified to
have the lowest $I_c$ in our structure \cite{Gerber96}.
The noise spectrum [c.f.~inset of
Fig.~\ref{figure3}(a)] is of Lorentzian-type which
indicates that it is dominated by a single strong
fluctuator. Moreover, we took images in the same
cooling field $B_0=6\,\rm\mu T$ which did not show
occupation of this particular site with a vortex. In
those cases the low frequency noise was low and
comparable to the values obtained for $B_0\le
4\,\rm\mu T$. This suggests strongly that the large
increase in low frequency flux noise at $B_0=6\,\rm\mu
T$ is due to the presence of this single fluctuating
vortex.

In conclusion, we demonstrated that LTSEM provides images of the
spatial distribution of vortices pinned in a high-$T_c$
dc SQUID multilayer magnetometer. The combination of
vortex imaging with low frequency noise measurements
enabled us to identify a single vortex being
responsible for a large excess low-frequency flux noise
in the device under test.  In general, the combination
of vortex imaging and low-frequency noise measurements
in variable magnetic field is expected to give new
insights into the mechanisms which are responsible for
generation of the frequently observed high levels of
flux noise in high-$T_c$ SQUIDs and magnetometers.

We thank John Clarke and G. Nichols for
providing detailed layouts of their SQUID-readout
electronics. The integrated magnetometer was fabricated at UC Berkeley.
We gratefully acknowledge collaboration with F. Ludwig, E.
Dantsker, A. H. Miklich, D. T. Nemeth and John Clarke
who provided this device. This work was supported by the
Deutsche Forschungsgemeinschaft (Hu251/27-1,Gr1132/11-1)
and the ESF VORTEX program.

\vspace*{-5mm}               %

\end{multicols}               %
\end{document}